\documentclass[seceq]{ptptex}

\usepackage{graphicx}
\title{%        %You can use \\ for explicit line-break
Five-quark picture of ${\rm \Lambda}(1405)$ in anisotropic lattice QCD
}

\author{%       %Use \scshape  for the family name
Noriyoshi \textsc{Ishii}$^{1}$\footnote{ e-mail address:
  ishii@ribf.riken.jp
},
Takumi \textsc{Doi}$^{2,3}$,
Makoto \textsc{Oka}$^{4}$,
and
Hideo \textsc{Suganuma}$^{5}$
}

\inst{%         %Affiliation, neglected when [addenda] or [errata]
$^1$ Center for Computational Sciences, Univ. of Tsukuba,
Tsukuba 305--8577, Japan\\
$^{2}$ Dept. of Physics and Astronomy, Univ. of Kentucky, Lexington KY 40506, USA\\
$^{3}$
RIKEN BNL Research Center, Brookhaven National Lab., New York 11973, USA\\
$^{4}$ Department of Physics, Tokyo Institute of Technology,
Tokyo 152--8551, Japan\\
$^{5}$ Department of Phyics, Kyoto University,
Kyoto 606--8502, Japan\\
}

\abst{%         %this abstract is neglected when [addenda] or [errata]
Five-quark (5Q) picture of  ${\rm \Lambda}(1405)$  is  studied using  quenched
lattice  QCD  with an  exotic  5Q  operator  of N$\bar{\rm K}$  type.   To
discreminate  mere  N$\bar{\rm K}$ and  ${\rm \Sigma}\pi$  scattering states,  
Hybrid Boundary Condition (HBC), a flavor-dependent boundary condition,  
is imposed on  the quark fields along  spatial direction.  
5Q  mass $m_{\rm 5Q}\simeq 1.89$  GeV is obtained  after the chiral  
extrapolation to  the physical  quark mass region, which is too heavy 
to be identified with ${\rm \Lambda}(1405)$.  
Then, ${\rm \Lambda}(1405)$ seems neither a pure 3Q state nor a pure 5Q state,
and therefore we present an interesting possibility that ${\rm \Lambda}(1405)$ 
is a mixed state of 3Q and 5Q states.}

\begin{document}

\maketitle

${\rm \Lambda}(1405)$ is an $I=0, S=-1,  Q=0$ negative-parity baryon.  As is
obvious from its  quantum number, it contains a  (heavy) strange quark
as a valence quark.  
Nevertheless, ${\rm \Lambda}(1405)$ is the lightest 
negative-parity baryon. (The lightest non-strange baryon is N$^*$(1520).)
In a quark-model interpretation, ${\rm \Lambda}(1405)$ is identified as 
a flavor SU(3) singlet baryon.
For this assignment, however, the anomalous value of the LS force is to be introduced.
Then, one may wonder if there  may be something behind, i.e., it may
involve some exotic structure.
Indeed,   it  has   been  conjectured,   for  a   long   period,  that
${\rm \Lambda}(1405)$  may be a  bound state  of N  and $\bar{\rm K}$  via the
strong interaction, i.e., a 5Q object  rather than a 3Q one.
%
%Note  that a Nambu-Goldstone mode  makes  it possible  for 5Q  state to  be
%lighter than p-wave 3Q state.
%
If this is the case, its binding energy is $m_N + m_K - 1405 \simeq 30$ MeV, 
which seems to be natural magnitude of the binding energy for the hadronic
molecule.

Baryon spectra in flavor SU(3)  sector was studied by quenched lattice
QCD   \cite{nemoto,melnitchouk,liu,burch}.   It  was   suggested  that
${\rm \Lambda}(1405)$  may have  an exceptional  feature.  For  instance, in
Ref.[\ref{nemoto}], it was reported that baryon masses in flavor SU(3)
sector can  be reproduced by quenched  lattice QCD within  about 10 \%
deviation  except   for  ${\rm \Lambda}(1405)$,  for   which  a  significant
overestimate of more than 300 MeV is reported.
One of the most attractive explanations for this is provided by the 5Q
picture of  ${\rm \Lambda}(1405)$, i.e., if ${\rm \Lambda}(1405)$  is dominated by
5Q  components,  it  should  be  difficult  for  quenched  QCD  with  a
conventional 3Q interpolating field to reproduce it.

In this paper, we will  consider the 5Q picture of ${\rm \Lambda}(1405)$ by
making  a constructive use  of quenched  QCD.  We  attempt to  use the
suppressed  contributions from  $q\bar{q}$  loops in  quenched QCD  in
order  to  obtain  an  additional  information on  a  possible  exotic
structure of ${\rm \Lambda}(1405)$ in the following way.
An ordinary  3Q interpolating field  fully couples to  3Q intermediate
states,  whereas it  couples  to  5Q intermediate  states  only in  an
imperfect manner.   On the  other hand, a  5Q interpolating  field can
fully couple  to 5Q intermediate  states, whereas its couplings  to 3Q
intermediate states  vanish at all,  once we neglect  the contribution
from  the  annihilation  diagram.   We  will use  these  quenched  QCD
features to study the nature  of ${\rm \Lambda}(1405)$ by comparing the mass
spectrum  obtained with an  ordinary 3Q  interpolating field  and that
with a 5Q interpolating field neglecting the annihilation diagram.
For 3Q  interpolating field, we employ a  flavor-singlet interpolating
field as
$
  {\rm \Lambda}^{\rm (3Q)}
  \equiv
  \epsilon_{abc}\left[
    (u_a^T C\gamma_5 d_b) s_c
    + (d_a^T C\gamma_5 s_b) u_c
    \right.
$
$
    \left.
   + (s_a^T C\gamma_5 u_b) d_c
    \right],
$
where  $a,b,c$ denote color  indices, and  $C\equiv \gamma_0\gamma_2$
denotes the charge conjugation  matrix. For 5Q interpolating field, we
employ an iso-scalar 5Q interpolating field of N$\bar{\rm K}$ type as
$
  {\rm \Lambda}^{\rm (5Q)}
  \equiv
  p {\rm K}^- - n \bar{\rm K}^0,
$
where  $p\equiv \epsilon_{abc}  (u_a^T C\gamma_5  d_b)  u_c$, $n\equiv
\epsilon_{abc}  (u_a^T  C\gamma_5   d_b)  d_c$,  K$^-  \equiv  \bar{u}
i\gamma_5   s$,  and  $\bar{\rm K}^0   \equiv  -\bar{d}i\gamma_5   s$.

5Q calculation involves an obstacle. Empirically, the mass gap between
${\rm \Lambda}(1405)$ and the N$\bar{\rm K}$ threshold is only about 30 MeV,
which is too small for a practical lattice QCD calculation to identify
these two states separately.
To  avoid this, we  adopt a  flavor-dependent boundary  condition (BC)
along    the    spatial    directions    (``{\it    Hybrid    Boundary
Condition(HBC)}'')  \cite{ishii-5Q,iida}, which  was  proposed in  the
studies of ${\rm \Theta}^+(1540)$.
HBC  consists of  anti-periodic BC  on $u$  and $d$  quark  fields and
periodic BC on $s$ quark field.
Since  ${\rm \Lambda}^{\rm (5Q)}$ field  contains  even number  of  $u$ and  $d$
fields, it is subject to the  periodic BC, which allows us to consider
the rest frame of the 5Q system ${\rm \Lambda}(1405)$.
In contrast, since $p$, $n$, K$^-$, $\bar{\rm K}^0$ fields contain odd
number of  $u$ and $d$ fields,  they are subject  to the anti-periodic
BC.   Their spatial  momenta are  discretized as  $\vec p  =  ((2n_x +
1)\pi/L,  (2n_y   +  1)\pi/L,  (2n_z  +  1)\pi/L)$,   where  $n_i  \in
\mathbb{Z}$, $L$ is  the spatial extension of the  lattice.  Note that
$|\vec  p|$ cannot  vanish with  HBC.   Its minimum  value is  $p_{\rm
min}\equiv \sqrt{3}\pi/L$, owing to  which N$\bar{\rm K}$ threshold is
raised from $E_{\rm  PBC,th} \simeq m_{N} + m_{K}$  for PBC to $E_{\rm
HBC, th}\simeq\sqrt{m_N^2 +  3\pi^2/L^2} +\sqrt{m_{K}^2 + 3\pi^2/L^2}$
for HBC.
For $L\simeq 2.2$ fm, the minimum value of $|\vec p|$ amounts to about
$499$ MeV leading to the shift  in the threshold of more than 200 MeV,
which make it possible to distinguish a possible ${\rm \Lambda}(1405)$ state
from N$\bar{\rm K}$ threshold.

The 5Q calculation still involves a difficulty.
${\rm \Lambda}(1405)$ is expected to be embedded in $\pi{\rm \Sigma}$ ``continuum''
even with HBC.
Note that ${\rm \Sigma}\pi$ threshold is not raised by HBC.
We  have to  distinguish ${\rm \Lambda}(1405)$  as a  compact 5Q  state from
$\pi{\rm \Sigma}$ scattering states.
If we were to impose such a spatial BC that the periodic BC is imposed
on $\bar{u}$, $\bar{d}$, and $\bar{s}$ fields, while the anti-periodic
BC is imposed on $u$, $d$ and $s$ fields, then the both the N$\bar{\rm K}$
and ${\rm \Sigma}\pi$ threshold could be raised more than 200 MeV.
However, this  is problematic,  since it does  not respect  the charge
conjugation symmetry.
Instead, we virtually introduce  additional flavors $u'$ and $d'$, and
regard $\bar{u}$  and $\bar{d}$  in 5Q ${\rm \Lambda}(1405)$  as anti-quarks
for  these two  additional flavors,  i.e., $\bar{u}'$  and $\bar{d}'$,
respectively.
We emphasize  that this will  not change anything, because  we neglect
the annihilation diagram as is mentioned before.
%
% curry
%
Now, we consider  a modified HBC, which will be  referred to as ``{\it
HBC2}''.  HBC2  consists of the periodic  BC on $u'$  and $d'$ fields,
and the anti-periodic BC on $u$, $d$, $s$ fields.
Note that ${\rm \Lambda}^{\rm (5Q)}$ field consists of four quarks with original
flavor  and  one anti-quark  with  additional  flavor.  N,  ${\rm \Sigma}$
fields consist  of three quarks  with original flavor.   $\bar{\rm K}$ and
$\pi$  fields  consist of  one  quark  with  original flavor  and  one
anti-quark   with   additional  flavor.    By   repeating  a   similar
consideration, we convince  ourselves that ${\rm \Lambda}^{\rm (5Q)}$ is subject
to the  periodic BC,  whereas N, ${\rm \Sigma}$,  $\bar{\rm K}$ and  $\pi$ are
subject to the anti-periodic BC.
In this  way, HBC2  can raise the  thresholds both for  N$\bar{\rm K}$ and
${\rm \Sigma}\pi$ while keeping a  possible compact 5Q ${\rm \Lambda}(1405)$ state
unaffected.

For precision measurements, we  use anisotropic lattice QCD, which has
4  times finer  mesh along  the  temporal direction  than the  spatial
directions as $a_s/a_t = 4$ \cite{nemoto,matufuru}.
We employ the standard Wilson  gauge action at $\beta=5.75$ and $O(a)$
improved  (clover)  quark  action  with  $\kappa=0.1210(0.010)0.1240$,
which covers  roughly the  quark mass region  of $m_s \le  m_{u,d} \le
2m_s$.
$\kappa_s=0.1240$ is fixed for $s$ quark, while $0.1210 \le \kappa \le
0.1240$  is  used for  chiral  extrapolation  for  $u$ and  $d$  quark
masses. The  lattice spacing is  determined with the  Sommer parameter
$r_0^{-1}=395$  MeV, which  leads to  the spatial  lattice  spacing of
$a_s^{-1}=1.10$ GeV ($a_s \simeq 0.18$  fm).
We use  the lattice  size $12^3\times 128$.  The spatial  extension is
$L\simeq 2.2$ fm.
We use totally 2000 gauge field configurations.
Along the  temporal direction, we impose  Dirichlet boundary condition
on $t=0$ plane.
We employ a Gaussian smeared source with the Gaussian size $\rho\simeq
0.4$ fm, which  is located on $t=t_0\equiv 64$  plane.  We utilise the
time-reversal and charge  conjugation symmetries to effectively double
the statistics.

\begin{figure}[tbh]
\begin{center}
\includegraphics[width=0.32\textwidth,angle=270]{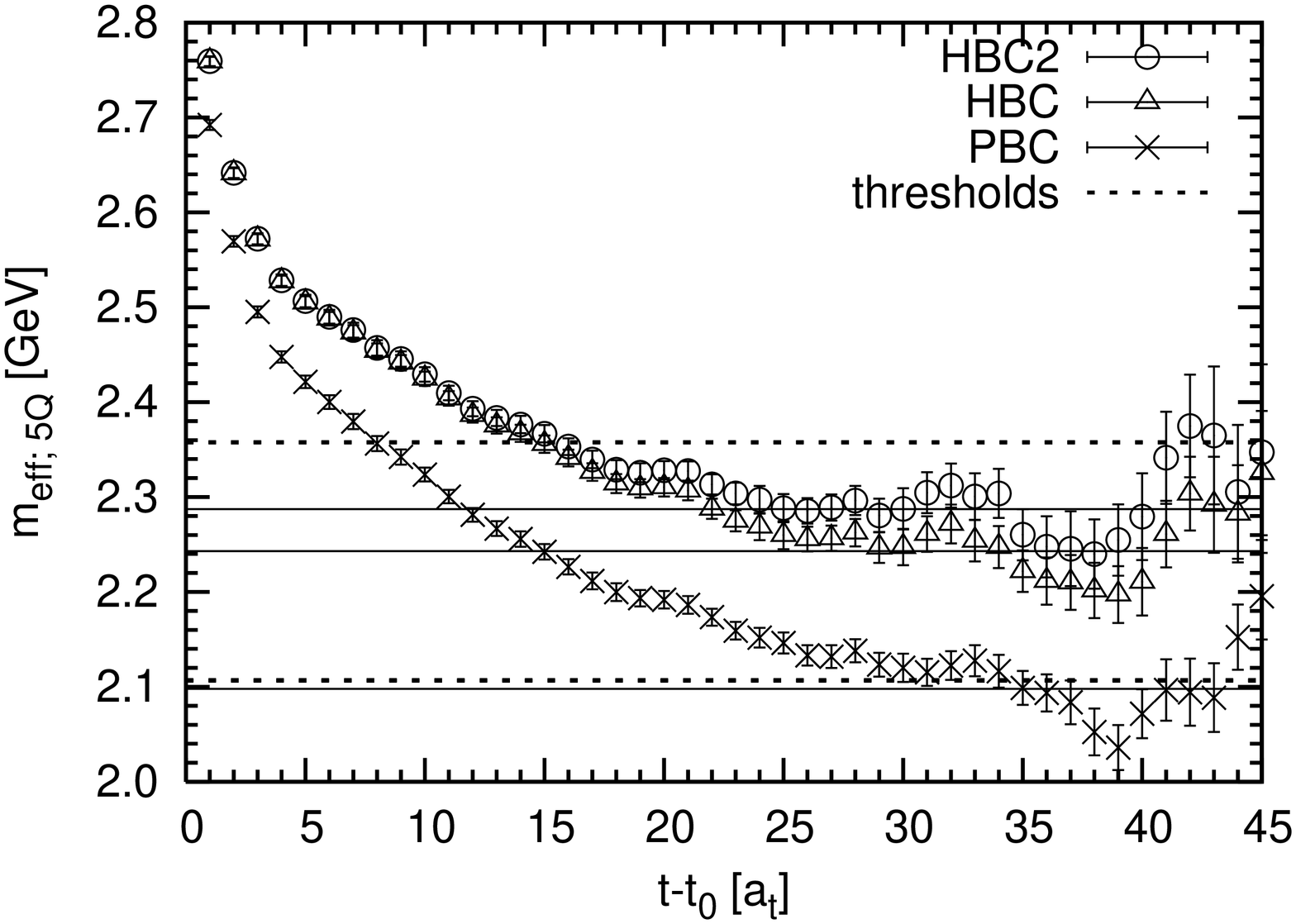}
\hspace*{1em}
\includegraphics[width=0.32\textwidth,angle=270]{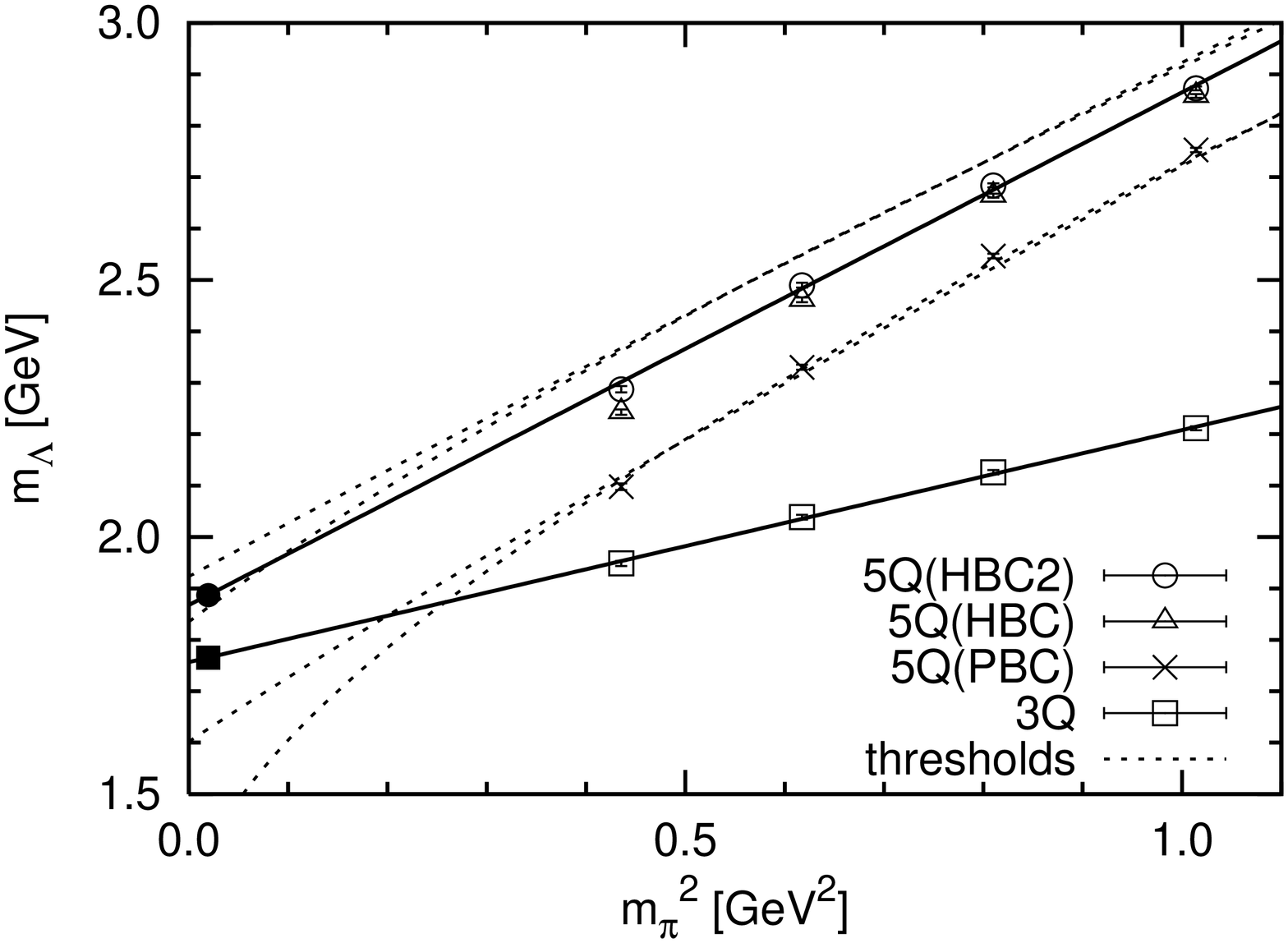}
\end{center}
\caption{The effective mass plot and the chiral extrapolation of 
the 5Q states which have same quantum number as the ${\rm \Lambda}$(1405).
Two types of the hybrid boundary condition (HBC, HBC2) are applied to 
raise the threshold.}
\end{figure}

Fig.~1(left)  shows  the  5Q  effective mass  plot  for  HBC2(circle),
HBC(triangle) and PBC(cross) for $\kappa=\kappa_{s}=0.1240$ case.
The lower  dotted line  denotes N$\bar{\rm K}$ and  ${\rm \Sigma}\pi$ thresholds
for PBC.
HBC raises N$\bar{\rm K}$ threshold, which  is denoted by the upper dotted
line.
HBC2 raises also  ${\rm \Sigma}\pi$ threshold, which is denoted  also by the
upper dotted line.
Note that, due  to the flavor SU(3) symmetry,  the (raised) N$\bar{\rm K}$
threshold coincides with the (raised) ${\rm \Sigma}\pi$ threshold.
We see that  a plateau is located at $t-t_0\in  [29,45]$ for PBC data.
It coincides with the PBC thresholds for N$\bar{\rm K}$ and ${\rm \Sigma}\pi$.
For HBC  and HBC2,  the plateaux  are raised above  by about  200 MeV,
which are located at $t \in [26,45]$.
It follows from these observations  that the plateau for PBC data does
not correspond to a compact 5Q state, but a scattering state.
%
%% Since the plateau  for HBC2 is located below  the raised thresholds by
%% about 60 MeV, it may be a  compact 5Q states. However, since 60 MeV is
%% not significantly large, it may  be caused by remaining interaction in
%% the finite volume.
%%
While the plateau  for HBC2 is located below  the raised thresholds by
small amount of $\sim 60$ MeV, this is most probably not an indication
of the existence of a  compact 5Q states, considering the energy shift
caused by remaining interaction in the finite volume.
To clarify which is the case, further investigation is necessary.
Since HBC  does not  affect ${\rm \Sigma}\pi$ threshold,  the fact  that the
plateau  for HBC  data is  raised by  about 200  MeV implies  that our
N$\bar{\rm K}$-type  interpolating field  has  only a  small overlap  with
${\rm \Sigma}\pi$ state.
The results  of single exponential  fits in these plateau  regions are
shown by solid lines.

Fig.~1(right) shows  the results  of linear chiral  extrapolations. 5Q
data for  HBC2, HBC,  and PBC are  denoted with circle,  triangle, and
cross,  respectively.  The  result of  the 3Q  data is  also  shown by
square.  Dotted curves denote  the raised and un-raised thresholds for
N$\bar{\rm K}$ and ${\rm \Sigma}\pi$.
The solid lines denote the results of the chiral extrapolations for 5Q
data with HBC2 and 3Q data.
Note that  the 5Q result  $m_{\rm 5Q}=1.887(9)$ GeV appears about  120 MeV
above the 3Q result $m_{\rm 3Q}=1.765(8)$  GeV, which is located about 360
MeV above the empirical value.
This  5Q  state is  too  heavy to  be  identified  with the  empirical
${\rm \Lambda}(1405)$, which  implies that ${\rm \Lambda}(1405)$ is not  a pure 5Q
state.

To  conclude, we  have studied  the 5Q  picture of  ${\rm \Lambda}(1405)$ by
using quenched anisotropic lattice QCD.
By taking an advantage of quenched  QCD, we have attempted to obtain a
key to investigate a possible exotic structure of ${\rm \Lambda}(1405)$.
We have  compared the results of  the 5Q interpolating  field with the
result  of the  3Q interpolating  field, and  have found  that  the 5Q
results appear  about 120 MeV  above the 3Q  result. Note that  the 3Q
result already overestimate the empirical  value 1405 MeV by about 360
MeV.
This has implied that ${\rm \Lambda}(1405)$  is not a 5Q dominant state. The
5Q  picture alone  does  not  provide a  solution  to the  significant
overestimate of the mass of ${\rm \Lambda}(1405)$ in quenched QCD.
{\it Since ${\rm \Lambda}(1405)$ is neither a pure 3Q state nor a pure 5Q state,
we present an interesting possibility that  ${\rm \Lambda}(1405)$ is a mixed
state of 3Q and 5Q states.}
%
%% The smaller energy  may be obtained in the larger  space of the states
%% where we seek for a solution.
(Note that, the energy is generally reduced, if one seeks for a solution in a larger space.)
The chiral quark effect may be also interesting \cite{liu}.
Although  quenched  QCD  with  3Q  interpolating  field  leads  to  an
imperfect  overlap  with   intermediate  5Q  states,  5Q  contribution
increases as the smaller quark mass region is approached.
Of course,  the annihilation  diagram may be  important, which  we have
neglected in our calculation to save the computational time.
Since ${\rm \Lambda}(1405)$ is such an interesting hadron, which may provide
us with  a possible  exotic structure, we  will keep studying  on this
interesting target from every aspect.

%% \begin{center}
%% Acknowledgements
%% \end{center}
\section*{Acknowledgements}
The authors acknowledge the Yukawa Institute for Theoretical Physics at Kyoto University 
for useful discussions during the YKIS2006 on ``New Frontiers in QCD''.
T.~D. is supported by Special Postdoctoral Research
Program of RIKEN and U.S. DOE grant DE-FG05-84ER40154.
Lattice QCD Monte  Carlo calculations have been done  with NEC SX-5 at
Osaka University. 

%%%%%%%%%%%%%%%%%%%%%%%%%%%%%%%%%%%%%%%%%%%%%%%%%%%%%%%%%%%%%
% Some macros are available for the bibliography:
%  o for general use
%    \JL : general journals                 \andvol : Vol (Year) Page
%  o for individual journal 
%    \AJ   : Astrophys. J.           \NC         : Nuovo Cim.
%    \ANN  : Ann. of Phys.           \NPA, \NPB  : Nucl. Phys. [A,B]
%    \CMP  : Commun. Math. Phys.     \PLA, \PLB  : Phys. Lett. [A,B]
%    \IJMP : Int. J. Mod. Phys.      \PRA - \PRE : Phys. Rev. [A-E]     
%    \JHEP : J. High Energy Phys.    \PRL        : Phys. Rev. Lett.
%    \JMP  : J. Math. Phys.          \PRP        : Phys. Rep.
%    \JP   : J. of Phys.             \PTP        : Prog. Theor. Phys.     
%    \JPSJ : J. Phys. Soc. Jpn.      \PTPS       : Prog. Theor. Phys. Suppl.
% Usage:
%  \PRD{45,1990,345}          ==> Phys.~Rev.\ \textbf{D45} (1990), 345
%  \JL{Nature,418,2002,123}   ==> Nature \textbf{418} (2002), 123
%  \andvol{B123,1995,1020}    ==> \textbf{B123} (1995), 1020
%%%%%%%%%%%%%%%%%%%%%%%%%%%%%%%%%%%%%%%%%%%%%%%%%%%%%%%%%%%%%

\end{document}